\documentclass[11pt]{article}
\usepackage{amsmath,amssymb,color,graphics,epsfig,cite}

\usepackage{amsfonts}

\newcommand{\hoch}[1]{$\, ^{#1}$}

\makeatletter
\@addtoreset{equation}{section}
\makeatother

\newcommand{\be}{\begin{equation}}
\newcommand{\ee}{\end{equation}}
\newcommand{\bea}{\setlength\arraycolsep{2pt} \begin{eqnarray}}
\newcommand{\eea}{\end{eqnarray}}
\newcommand{\nn}{\nonumber}

\def\crampest{\medmuskip = 1mu plus 1mu minus 1mu}

\def\ft#1#2{{\textstyle{\frac{\scriptstyle #1}{\scriptstyle #2} } }}
\def\fft#1#2{{\frac{#1}{#2}}}

\def\0{{\sst{(0)}}}
\def\1{{\sst{(1)}}}
\def\2{{\sst{(2)}}}
\def\3{{\sst{(3)}}}
\def\4{{\sst{(4)}}}
\def\5{{\sst{(5)}}}
\def\6{{\sst{(6)}}}
\def\7{{\sst{(7)}}}
\def\8{{\sst{(8)}}}
\def\sst#1{{\scriptscriptstyle #1}}

\def\del{{\partial}}

\def\cH{{{\cal H}}}

\begin{document}

\begin{flushright}
\hfill{MI-TH-1537}

\end{flushright}

\begin{center}
{\Large {\bf Thermodynamics of Charged Black Holes \\
in Einstein-Horndeski-Maxwell Theory}}

\vspace{15pt}
{\bf Xing-Hui Feng\hoch{1}, Hai-Shan Liu\hoch{2,3}, H. L\"u\hoch{1} and C.N. Pope\hoch{3,4}}

\vspace{10pt}

\hoch{1}{\it Center for Advanced Quantum Studies, Department of Physics, \\
Beijing Normal University, Beijing 100875, China}

\vspace{10pt}

\hoch{2} {\it Institute for Advanced Physics \& Mathematics,\\
Zhejiang University of Technology, Hangzhou 310023, China}

\vspace{10pt}

\hoch{3} {\it George P. \& Cynthia Woods Mitchell  Institute
for Fundamental Physics and Astronomy,\\
Texas A\&M University, College Station, TX 77843, USA}

\vspace{10pt}

\hoch{4}{\it DAMTP, Centre for Mathematical Sciences,
 Cambridge University,\\  Wilberforce Road, Cambridge CB3 OWA, UK}

\vspace{20pt}

\underline{ABSTRACT}
\end{center}

    We extend an earlier investigation of the thermodynamics of
static black holes in an Einstein-Horndeski theory of gravity coupled to a scalar
field, by including now an electromagnetic field as well.  By studying the
two-parameter families of charged static black
holes, we obtain much more powerful constraints on the thermodynamics
since, unlike in the uncharged one-parameter case, now the right-hand
side of the first law is not automatically integrable.  In fact, this allows us
to demonstrate that
there must be an additional contribution in the first law, over and above
the usual terms expected for charged black holes.  The origin of the
extra contribution can be attributed to the behaviour of the scalar field
on the horizon of the black hole. We carry out the calculations in four
dimensions and also in general dimensions.  We also derive
the ratio of viscosity to entropy for the dual boundary field theory,
showing that the usual viscosity bound for isotropic solutions can be violated, with the ratio depending on the mass and charge.

\vfill {\footnotesize xhfengp@mail.bnu.edu.cn \ \ \  hsliu.zju@gmail.com \ \ \ mrhonglu@gmail.com\ \ \
pope@physics.tamu.edu}

\thispagestyle{empty}

\pagebreak

\tableofcontents
\addtocontents{toc}{\protect\setcounter{tocdepth}{2}}



\section{Introduction}

Black holes are the most fundamental objects predicted by Einstein's
theory of gravity.  They have been extensively studied, and many
important properties have been established.  Subject to certain
assumptions, the no-hair theorems establish that all properties of a black hole are completely characterized by a
few conserved charges (see, for example, \cite{nohair}). However,
general arguments and numerical evidence suggests that
black holes in Einstein-scalar theories can develop scalar ``hair,''
both in asymptotically-AdS and
asymptotically-flat spacetimes \cite{Sudarsky:2002mk, Nucamendi:1995ex}.
Recently, many explicit examples of black holes carrying scalar hair have
been found, in four and higher
dimensions \cite{Anabalon:2012ta, Anabalon:2013qua, Anabalon:2013sra, Gonzalez:2013aca, Acena:2013jya, Feng:2013tza,herdeiro,Wen:2015xea,Fan:2015tua,Fan:2015oca
}. It was shown that the first law of
thermodynamics of
(charged) AdS black holes can be modified by the boundary conditions of
the scalar field at asymptotic infinity \cite{llscalar,Lu:2014maa}. The
first such explicit example, with a closed-form expression for the solution,
was provided by the Kaluza-Klein AdS dyonic black hole \cite{Lu:2013ura}.
It is important to emphasise that we take
the first law, more or less by definition, to mean the relation
between the infinitesimal variations of the parameters characterising the
asymptotic form of the black hole solution near infinity and the variation
of the parameters characterising the near-horizon expression
for the black hole solution.  In a simple example such as a
Reissner-Nordstr\"om (RN) black hole of mass $M$ and charge $Q$, this
relation equates $\delta M-\Phi\, \delta Q$ at infinity (in the gauge
where $\Phi=0$ on the horizon) to $T\delta S$
on the horizon.

In a typical theory of Einstein gravity, matter fields couple to
gravity minimally through the metric. However, theories such as
Brans-Dicke theory \cite{Brans:1961sx}, which involves a
non-minimally coupled scalar, have also been widely studied.
Recently, Galileon theories \cite{Nicolis:2008in} have attracted attention,
in which
a scalar field has non-minimal derivative couplings to the curvature
tensor. In fact, general classes of theories of this kind were
constructed  by Horndeski in the early 1970s \cite{Horndeski:1974wa}.
These gravity-scalar theories are characterised by the property that
both the gravity and the scalar field equations contain
no more than second derivatives, which is analogous to
the situation in Lovelock gravities \cite{Lovelock:1971yv}.

     The thermodynamics of black holes in Einstein-Horndeski gravity was
studied
in \cite{unchargedhorndeski}. The theory studied there contains the usual
Schwarzschild black hole, whose thermodynamics is the same as
in Einstein gravity.  A no-go theorem was established for asymptotically-flat
spacetimes, showing that no further black hole solutions existed
\cite{huinic,sotzho1}.  When the theory contains a cosmological constant,
there is in addition a branch of
scalar-hairy black holes \cite{aco}. (See also
\cite{Rinaldi:2012vy,Babichev:2013cya,Minamitsuji:2013ura,sotzho2,babi2}.)
Surprises
emerged for these black hole solutions \cite{unchargedhorndeski}.
Firstly, the standard Wald entropy formula \cite{wald1,wald2}, namely
\be
S_{\sst W}=-\fft{1}{8} \int_+ d^{n-2} x \sqrt{h}\,
\fft{\partial L}{\partial R^{abcd}}
\epsilon^{ab} \epsilon^{cd}\,,\label{waldentropy}
\ee
where $L$ is defined by the action $I=\int d^n x \sqrt{-g} L$, is no
longer valid.  Secondly, the quantum statistical relation
\cite{Gibbons:1976ue} appears to be no longer valid, and possibly this
may be related to the theory's not admitting a Hamiltonian formulation.

   To resolve the issue and derive a proper first law of black hole
thermodynamics, the Wald procedure for computing a conserved Hamiltonian
variation was applied to the black holes in \cite{unchargedhorndeski},
paying particular attention to details of the scalar contribution. It turns
out that the first law for the black holes in this theory is modified
as a consequence of the unusual behavior of the scalar field on the event
horizon.  It
arises because the derivative of the scalar field $\chi$ diverges
on the horizon, although there is no physical divergence since all invariants,
including $g^{\mu\nu}\, \del_\mu\chi\, \del_\nu\chi$, remain finite.

The first law of black hole thermodynamics is a consequence of the
equality between the variation of the Hamiltonian at asymptotic infinity
and on the horizon, i.e.
\be
\delta {\cal H}_\infty=\delta {\cal H}_+\,.\label{waldid}
\ee
This identity was shown to be valid for the scalar-hairy black holes in
Einstein-Horndeski gravity \cite{unchargedhorndeski}.  In order to be
able to interpret (\ref{waldid}) as a first law, one then needs to be able
to express the terms appearing in the variations on each side of the
equation in terms of thermodynamic variables characterising the solutions.
For the black hole solutions in \cite{unchargedhorndeski}, one can then
identify $\delta {\cal H}_{\infty}= \delta M$ and $\delta {\cal H}_+=
T\delta S$, and the first law $\delta M=T\delta S$ is thus ostensibly derived (with
$S$ effectively being {\it defined} by $\delta S=T^{-1}\, \delta{\cal H}_+$).
The weak point in this argument is that the static scalar-hairy
black holes studied in \cite{unchargedhorndeski}
depend on only one parameter, and hence there is no non-trivial
integrability check in the above derivation.  Furthermore, the
spherically-symmetric solutions are asymptotically locally AdS, and hence
it is difficult to find any independent method to calculate the mass.
Thus in this first law, only $T$ can be unambiguously calculated
independently of the
Wald procedure, and it follows that the validity of $dM=TdS$ for the
scalar-hairy black hole in \cite{unchargedhorndeski} remains questionable,
as was discussed in that paper.

In this paper, we consider Einstein-Horndeski gravity minimally coupled to
a Maxwell field.  As in the case of the two-parameter RN black holes
of pure Einstein-Maxwell theory,
where the first
law $dM=TdS + \Phi_e dQ_e$
implies that there is a non-trivial integrability condition
$dT\wedge dS + d\Phi_e\wedge dQ_e=0$, the Einstein-Horndeski-Maxwell (EHM)
theory we shall consider admits two-parameter static black hole
solutions, and so again we can obtain further non-trivial insights into the
form of the first law by means of integrability considerations.
The static electrically-charged black hole solutions in EHM gravity were
obtained in \cite{ac}. (See also \cite{Erices:2015xua}.) We shall use
the Wald formalism to derive the first law.  For a minimally-coupled
Maxwell field, the electric charge $Q_e$ and the electrostatic potential
$\Phi_e$ (i.e the potential difference between the horizon and
asymptotic infinity) can be independently and unambiguously defined.
 We find that for the charged black holes, the Wald identity
(\ref{waldid}) is indeed always satisfied, but reading off
a first law from this is somewhat subtle.
In a gauge choice where the gauge potential $A$ vanishes on the horizon,
we find that $\delta {\cal H}_\infty = \delta M - \Phi_e \delta Q_e$, as
one would have expected.  However, at the horizon
we find $\delta {\cal H}_+\ne T \delta S$, and in fact
$T^{-1}\delta {\cal H}_+$ is not integrable.  In order to
complete the first law, we find it necessary to
introduce the concept of a global scalar charge, which can give a
non-trivial contribution to $\delta {\cal H}$ on the horizon.

The paper is organised as follows.  In section 2 we introduce the
charged
Horndeski theory that we shall be considering, and present the equations
of motion for a static black-hole ansatz. We then review three kinds of
static black hole solutions in four dimensions in section 3.  Next,
we analyze the thermodynamics of these black holes in section 4, by using
the Wald formalism. We find that the entropy has a
standard contribution of one quarter the area of the event horizon
(in accordance with the Wald entropy formula) but with an
additional  contribution from the scalar field.  Furthermore,
the first law also requires an additional contribution from the
scalar field. In section 5, we generalize these four-dimensional
results to arbitrary dimensions. For planar AdS black holes, there
exists a global scaling symmetry.  We study the corresponding Noether
charge and obtain the generalized Smarr relation in section 6.
In section 7, the ratio of viscosity to  entropy density is calculated.
The paper ends with conclusions in
section 8.

\section{The Theory and Equations of Motion}

In this paper we consider an Einstein-Horndeski gravity, minimally
coupled to a Maxwell field, with the action given by
\bea
I = \fft{1}{16\pi}\int d^nx\sqrt{-g}\, L\,, \quad
L = \kappa(R-2\Lambda - \ft14 F^2)-
\ft{1}{2}(\alpha g^{\mu\nu}-\gamma G^{\mu\nu})
\, \del_\mu\chi\, \del_\nu\chi\,,\label{action}
\eea
where $\kappa$, $\alpha$ and $\gamma$ are coupling constants,
$G_{\mu\nu} \equiv R_{\mu\nu} - \fft12 R g_{\mu\nu}$ is the Einstein tensor,
and $F=dA$ is the electromagnetic field strength.  The theory consists of the
metric $g_{\mu\nu}$, the $U(1)$ gauge potential $A=A_\mu dx^\mu$ and
the scalar $\chi$.  The scalar is axionic, appearing in the Lagrangian
only through its derivative.  Thus the Lagrangian is invariant under the
constant shift transformation $\chi\rightarrow \chi + c$.  One may gauge
this symmetry and obtain Einstein-vector gravity with a non-minimally
coupled vector field \cite{Geng:2015kvs}, but we shall not consider this
possibility in the present paper.

The equations of motion following from the variations
of the metric $g_{\mu\nu}$, the scalar $\chi$ and the Maxwell
potential $A=A_\mu dx^\mu$ are given by
\bea
E_{\mu\nu} &\equiv& \kappa (G_{\mu\nu} +\Lambda g_{\mu\nu} - \fft12 F_{\mu\nu}^2 + \fft18 F^2 g_{\mu\nu})  \cr
&& -\ft12\alpha \Big(\partial_\mu \chi \partial_\nu \chi - \ft12 g_{\mu\nu} (\partial\chi)^2\Big)-\ft12\gamma \Big(\ft12\partial_\mu\chi \partial_\nu \chi R - 2\partial_\rho
\chi\, \partial_{(\mu}\chi\, R_{\nu)}{}^\rho \cr
&&- \partial_\rho\chi\partial_\sigma\chi\, R_{\mu}{}^\rho{}_\nu{}^\sigma -
(\nabla_\mu\nabla^\rho\chi)(\nabla_\nu\nabla_\rho\chi)+(\nabla_\mu\nabla_\nu\chi)
\Box\chi + \ft12 G_{\mu\nu} (\partial\chi)^2\cr
&&-g_{\mu\nu}\big[-\ft12(\nabla^\rho\nabla^\sigma\chi)
(\nabla_\rho\nabla_\sigma\chi) + \ft12(\Box\chi)^2 -
  \partial_\rho\chi\partial_\sigma\chi\,R^{\rho\sigma}\big]\Big)  = 0\,,\cr
E_\chi &\equiv&\nabla_\mu \big( (\alpha g^{\mu\nu} - \gamma G^{\mu\nu}) \nabla_\nu\chi\big) = 0\,, \qquad
E_A^\mu\equiv \nabla_\nu F^{\nu\mu}  =  0\,.\label{Heoms}
\eea
Although the theory involves a total of four derivatives, in the
equations of motion no field carries more than two derivatives, and
the total of four derivatives arise through nonlinearities. This is one of
the general
characteristic properties of Horndeski theories, and in consequence
the perturbations of any background are described by linear differential
equations of only second order in derivatives, and hence they can
be ghost free.

   In this paper, we are concerned with the properties of static charged
black hole solutions.  The most general static ansatz, after coordinate
gauge fixing, can be taken to be
\bea
ds^2 = - h(r) dr^2 + \fft{dr^2}{f(r)} + r^2 d\Omega^2_{\sst{n-2\,,
\epsilon}} \,, \quad \chi = \chi(r) \,, \quad A=\phi(r) \,dt\,, \label{staans}
\eea
where $d\Omega^2_{\sst{n-2\,, \epsilon}}$, with $\epsilon = 1\,, 0\,, -1$,
is the metric for the unit $S^{n-2}$, the $(n-2)$-torus or the unit
hyperbolic $(n-2)$-space.
The electrostatic potential $\phi$ can be expressed in terms of
the metric fields by using the Maxwell equation of motion $E_A^\mu = 0$,
which implies
\be
\phi' = \fft{q}{r^{n-2} }  \sqrt {\fft h f}  \,, \label{phieq}
\ee
where hereafter, we use a prime to denote a derivative with
respect to $r$.
The scalar equation of motion $E_\chi = 0$ gives
\be
\Big(r^{n-4} \sqrt{\fft{f}{h}}
\Big( \gamma \big( (n-2) r f h' + (n-2)(n-3) (f-\epsilon) h\big)
    -2\alpha r^2 h \Big)\chi'\Big)'=0\,.\label{scalareom}
\ee
There are two further independent equations that follow from
$E_{\mu\nu} = 0$, namely
\bea
&&4\kappa \Big((n-2)r f' + (n-2)(n-3) (f-\epsilon) +
2\Lambda r^2\Big) + 2 \kappa q^2 r^{6-2 n} + 2\alpha r^2 f\chi'^2\cr
&&\qquad+\gamma (n-2) \Big(4r f \chi'' + \big( 3r f' +
   (n-3)(f+\epsilon)\big)\chi' \Big)f \chi'
=0\,,\cr
&&  4\kappa \Big((n-2) r f h' + (n-2)(n-3) h (f-\epsilon) + 2\Lambda
    r^2 h\Big) + 2 \kappa q^2  r^{6-2 n} h\cr
&&\qquad -2 \alpha r^2 f h \chi'^2 + \gamma (n-2) \Big( 3 r f h' + (n-3) (3 f - \epsilon)h\Big) f \chi'^2
=0\,.
\eea

\section{Static Black Hole Solutions in Four Dimensions}

\label{sec:d4bh}

In the previous section, we obtained the equations for the static and
charged ansatz. The most general analytical solutions of these equations
would be difficult to obtain explicitly.
In \cite{ac}, a special class of solutions was obtained, by solving the
scalar equation (\ref{scalareom}) by taking
\be
\gamma \big( (n-2) r f h' + (n-2)(n-3) (f-\epsilon) h\big)
   -2\alpha r^2 h=0\,.
\label{specialeom}
\ee
The resulting solutions turn out to describe black holes.  In
\cite{unchargedhorndeski}, it was shown
that all the static neutral black hole solutions must
satisfy (\ref{specialeom}).
We expect the same to be true for the charged black holes.  In this section, we review the black holes constructed in \cite{ac}.

\bigskip
\noindent{\underline{\bf Asymptotically Minkowski solution}:}
\medskip

When $\alpha = 0$ and $\Lambda = 0$, the theory admits an
asymptotically Minkowski solution in four dimensions:
\bea
\chi'&=&\sqrt{-\fft{\kappa q^2}{2\gamma r^2\, f}}\,,\qquad
\phi =\phi_0- \fft{q}{r} + \fft{q^3}{24 r^3}\,,\qquad
f = \fft{64 r^4}{(8r^2 - q^2)^2}\,h\,,\cr
h&=&1 - \fft{\mu}{r} + \fft{q^2}{4r^2} - \fft{q^4}{192 r^4}\,. \label{minso}
\eea
The solution has three integration parameters, $\phi_0 \,,\, q$ and $\mu$,
of which two correspond to non-trivial parameters.
There are two curvature singularities; one is as usual at $r=0$, whilst
the other, where $f$ diverges,
is at $r=r_* = q/\sqrt 8$.  Here, we assume without loss of
generality that $q\ge 0$.  For the solution to describe a black hole, we
must have an event horizon $r=r_0$, which is the largest root of $h=0$.
Furthermore, we must require that $r_*$ be inside the event horizon.  This
implies that
\be
h(r_*) = \fft83 - \fft{\sqrt 8 \mu}{q} <0 \,,\qquad \Longrightarrow\qquad
\mu > \fft{2 \sqrt2 }{3} q \,.
\ee
With this condition satisfied, the functions $h$ and $f$ both run from 0 to
1 as $r$ runs from the horizon at $r=r_0$ to $r=\infty$. The Hawking
temperature can then be calculated in the standard way, leading to
\be
T = \fft{8 r_0^2 - q^2}{32 \pi r_0^3} \,.
\ee
Interestingly, the requirement that the singularity $r_* = q/\sqrt 8$ be
inside the event horizon implies that $T>0$. The temperature $T$ can
approach $0$, but can never reach $0$. This behavior is different from that
of the well-known RN black hole, and in fact is more in line with
the behaviour of conventional systems with respect to the third law of
thermodynamics.

It is worth remarking that these asymptotically-flat black holes provide
counterexamples to the no-go theorem of \cite{huinic,sotzho1}, in consequence
of the presence of the Maxwell field.

\bigskip
\noindent{\underline{\bf Asymptotically AdS solution}:}
\medskip

For generic  $\alpha \ne 0$ and $\Lambda \ne 0$, there exist asymptotically
(locally) AdS solutions. It is convenient to introduce two
parameters $(g \,, \beta)$ in place of the original two
parameters $(\alpha \,, \Lambda)$ in the Lagrangian, defined by
\be
\alpha=\ft12 (n-1)(n-2) g^2 \gamma\,,\qquad
\Lambda = -\ft12 (n-1)(n-2)g^2 \Big(1 + \fft{\beta\gamma}{2\kappa}\Big)\,.
\label{alphabeta}
\ee

   First, we present the simpler case of planar black holes,
which arise when $\epsilon = 0$.
The four-dimensional solution is then given by
\bea
\phi&=&\phi_0-\fft{q}{r} + \fft{\kappa q^3}{30 g^2(4\kappa + \beta \gamma)\,r^5}\,,\cr
\chi' &=& \sqrt{\beta - \fft{\kappa q^2}{6\gamma g^2 r^4}}\,\fft{1}{\sqrt{f}}\,,\qquad
f= \fft{36(4\kappa + \beta\gamma)^2 g^4r^8}{\big(\kappa q^2 -6 (4\kappa + \beta\gamma) g^2 r^4\big)^2}\, h\,,\cr
h&=& g^2 r^2 - \fft{\mu}{r} + \fft{\kappa q^2}{(4\kappa + \beta\gamma)r^2} -
\fft{\kappa^2 q^4}{60(4\kappa + \beta\gamma)^2g^2 r^6}\,. \label{ads0}
\eea
As in the previous asymptotically-Minkowski case, there is an
additional curvature singularity, as well as the usual one at $r=0$,
located now at
\be
r_* =\Big (\frac{\kappa  q^2}{6 g^2 (\beta  \gamma +4 \kappa )}
   \Big)^\fft14 \,,
\ee
where $f$ diverges. The requirement that this singularity be inside
the event horizon $r=r_0$ implies
\be
\mu >  \frac{16\times 2^\fft14 g^\fft12 \kappa ^{3/4} q^{3/2}}{
5\times 3^{3/4} (\beta  \gamma +4 \kappa )^{3/4}} \,.
\ee
Provided this condition is satisfied, the solutions describe asymptotically-AdS
planar black holes. The Hawking temperature is given by
\be
T = -\frac{\kappa  q^2-6 g^2 {r_0}^4 (\beta  \gamma +4 \kappa )}{8
\pi  {r_0}^3 (\beta  \gamma +4 \kappa )} \,.
\ee
As can be easily seen, the requirement that the singularity be
inside the black hole event horizon $r_*<r_0$ again leads to
the conclusion that $T>0$, and that it can never reach $0$.

Secondly, we consider $\epsilon = 1$, which yields static and
spherically-symmetric
black hole solutions, given by
\bea
\phi &=& \phi_0 -\frac{\kappa  q \left(3 g^2 q^2+8\right)}{2 r (\beta  \gamma +4 \kappa )} +\frac{\kappa  q^3}{6 r^3 (\beta  \gamma +4 \kappa )} +\frac{\sqrt{3} g q \, \arctan\left(\frac{1}{\sqrt{3} g r}\right) \left(3 g^2 \kappa  q^2-2 \beta  \gamma \right)}{2( \beta  \gamma +4 \kappa) } \,,\cr
\chi' &=& \sqrt{\fft{6\beta g^2\gamma r^4-\kappa q^2}{2\gamma r^2 (1 + 3g^2 r^2)}}\,\fft{1}{\sqrt{f}}\,,\qquad
f=\fft{4(4\kappa+\beta\gamma)^2 (1 + 3g^2 r^2)^2 r^4}{\big(
6(4\kappa +\beta\gamma) g^2 r^4 + \kappa(8r^2 - q^2)\big)^2}\,h\,,\cr
h&=& g^2 r^2 +\fft{4\kappa -\beta\gamma}{4\kappa + \beta\gamma} - \fft{\mu}{r}
+\fft{\kappa^2 q^2(3g^2 q^2 + 16)}{4(4\kappa + \beta\gamma)^2r^2} -
\fft{\kappa^2 q^4}{12(4\kappa + \beta\gamma ) ^2 r^4}\cr
&&-\fft{(2\beta\gamma - 3\kappa g^2 q^2)^2 \arctan(\fft1{\sqrt3\,g r})}{4 \sqrt{3} (4 \kappa +\beta\gamma)^2 g r} \label{ads1}
\eea
Besides the usual curvature singularity at the origin, there is
again a further curvature singularity at $r=r_*$, which is now given by
\be
6 g^2 r_*^4 (\beta  \gamma +4 \kappa )-\kappa  \left(q^2-8 r_*^2\right) =0 \,.
\ee
For the solutions to describe black holes, we require that the
largest root $r^*$ be less than $r_0$, the radius of the event horizon.
The Hawking temperature is
\be
T = \frac{6 g^2 r_0^{4} (4 \kappa +\beta  \gamma )+\kappa  \left(8 r_0^2-q^2 \right)}{8 \pi  r_0^3 (4 \kappa +\beta  \gamma)} \,.
\ee
As in the previous case, the requirement that the singularity $r_*$ should
be inside the event horizon implies that $T>0$. The requirement also sets
a constraint between mass parameter $\mu$ and charge parameter $q$.  Since the
explicit expression is rather complicated, we shall not present it here.

 \section{Thermodynamics}

Having described the charged black holes and their Hawking temperatures in
the previous section, we now turn to a study of their thermodynamic
properties, and in particular, we shall obtain the first law of black hole
thermodynamics that they obey.  The thermodynamics of the neutral
black holes in Horndeski gravity
was studied in \cite{unchargedhorndeski}.   It was shown there that the
standard Wald entropy formula requires modification in Einstein-Horndeski
gravity.  The reason for this is that the axionic scalar $\chi$ has a
branch-cut singularity at the horizon, with the associated consequence
the radial vielbein component of its derivative $\del_a\chi=
E_a^\mu\partial_\mu\chi$
approaches a constant, rather than vanishing, on the horizon.  This leads
to an additional contribution in the expression for the entropy.

   The neutral static black holes discussed in \cite{unchargedhorndeski}
depends on only one non-trivial parameter, and hence one can
always find a mass functional $M$ such that a first law of the
form $dM=TdS$ ostensibly holds, since $T dS$ is necessarily integrable.
The situation is very different now for the case of charged static black holes,
which depend on two non-trivial parameters, since integrability is no longer
{\it a priori} guaranteed in the two-dimensional parameter space of
solutions.

Although the Wald entropy formula acquires an additional scalar
contribution, the Wald conserved Hamiltonian formalism still works
for the neutral black holes, as analyzed in \cite{unchargedhorndeski}.
The subtlety lies in how to read off the first law of black hole
thermodynamics from the Wald formalism.  For this reason, we shall
also employ the Wald formalism to study the charged black holes. We first
give a brief review of the formalism, and then we apply it
to the charged black holes discussed in the previous section.

\subsection{Wald formalism}

Wald has developed a procedure that can be used to derive the first law
of thermodynamics,
by calculating the variation of a Hamiltonian associated with a
conserved Noether current \cite{wald1, wald2}. The Wald formalism has
been applied to the study of the first law of thermodynamics for
asymptotically-AdS black holes in a variety of theories, including
Einstein-scalar \cite{llscalar,Lu:2014maa},
Einstein-Proca \cite{Liu:2014tra}, Einstein-Yang-Mills \cite{Fan:2014ixa},
in gravities extended with quadratic-curvature invariants \cite{Fan:2014ala},
and also for Lifshitz black holes \cite{Liu:2014dva}. Starting from a
Lagrangian ${\cal L}$, its variation under a general variation of the
fields can be written as
\be
\delta { \cal L} = \text{e.o.m.} \, + \, \sqrt{-g} \, \nabla_\mu J^\mu \,,
\ee
where e.o.m. denotes terms proportional to the equations of motion for
the fields. From this one can define a 1-form $J_\1 = J_\mu dx^\mu$ and
its Hodge dual $\Theta_{\sst{(n-1)}} = (-1)^{n+1} * J\1$.
The next step  is to specialise to a variation that is induced by an
infinitesimal
diffeomorphism $\delta x^\mu = \xi^\mu$. One can show that
\be
J_{\sst{(n-1)}} \equiv \Theta_{\sst{(n-1)}} -
   i_\xi \,{* {\cal L}} = \text{e.o.m} - d\, {* J}_\2 \,,
\ee
where $i_\xi$ denotes a contraction of $\xi^\mu$ on the first index of
the $n$-form ${*{\cal L}}$. One can thus define an $(n-2)$-form $Q_{\sst{(n-2)}}
\equiv *J_\2$, such that $J_{\sst{(n-1)}} = dQ_{\sst{(n-2)}}$. Note that
we use the subscript notation $\sst{(p)}$ to denote a $p$-form. To make
contact with the first law of black hole thermodynamics, we take
$\xi^\mu$ to be the  Killing vector that becomes null on the horizon.
Wald shows that the variation of the Hamiltonian with respect to the
integration constants of a specific solution is given by
\be
\delta {\cal H}  = \fft{1}{16 \pi} \delta \int _c  J_{\sst{(n-1)}} -
\int_c d \big( i_\xi \Theta_{\sst{(n-1)}} \big) =
\fft{1}{16 \pi} \int_{\Sigma_{\sst{(n-2)}}}
\big( \delta Q - i_\xi \Theta_{\sst{(n-1)}}\big) \,,
\ee
where $c$ denotes an $(n-1)$-dimensional
Cauchy surface and $\Sigma_{\sst{(n-2)}}$ is its
boundary, which has two components, one at infinity and one on the horizon.
Thus according to the Wald formalism, the first law of black hole
thermodynamics is a consequence of the identity (\ref{waldid}),
which follows from the fact that $\delta Q-i_\xi\Theta_{\sst{(n-1)}}$
is an exact form.
It is worth commenting that the identity is valid as long as the spacetime
is well behaved in the region between the two surfaces.
It applies also for solitons, which have no horizon. In this
case it therefore implies $\delta {\cal H}_\infty=0$.
As we shall see in the next subsection, the Wald identity is satisfied by
all the solutions  presented in section \ref{sec:d4bh}, which is consistent
with the fact that these are all well-defined black holes.
As we shall see, there are subtleties involved in identifying
appropriate thermodynamical quantities that are compatible with
interpreting the Wald identity as a first law.

\subsection{Results for charged static black holes}

The Wald formalism for the general EHM theory (\ref{action}) is rather
involved, and we shall present here only the key results for the
static ansatz (\ref{staans}). (A detailed discussion of the Wald formalism
for Einstein-Horndeski gravity can be found in \cite{unchargedhorndeski}.) We find
\bea
(\delta Q -i_\xi \Theta)_{\text{min}}& =& -  r^{n-2}  \sqrt{\frac{h}{f}} \, \Big[ \, \kappa\,  \Big(\frac{n-2}{r}\delta f + \fft{f}{h}\, \phi\delta\phi' + \fft{\phi \phi'}{2} (\fft{\delta f}{h} - \fft{f \delta h}{h^2}) \Big) + \alpha f \chi' \delta \chi \, \Big] \,\Omega_{(n-2)}\,,\label{waldmin}\cr
(\delta Q -i_\xi \Theta)_{\text{non}} &=&  \ft{1}2(n-2)\gamma\, r^{n-3} \sqrt{\frac{h}{f}} f^2 \Big[-  \ft32 \chi'^2 \fft{\delta f}{f} - \delta (\chi'^2) \cr
&&\qquad\qquad\qquad\qquad\qquad+  \Big(\fft{n-3}{r} \,(1- \fft{\epsilon}{f})+
  \fft{h'}{h}\Big) \chi'\delta \chi \Big] \, \Omega_{(n-2)}\,,\label{waldgamma}
\eea
where the subscript ``min'' or ``non'' indicates terms that are related
to the minimally-coupled part or the non-minimally coupled part of the
action (\ref{action}) respectively.  We find that the coefficient of the
term $\chi' \delta \chi $ is proportional to the constraint equation
(\ref{specialeom}), and so for the solutions we are
considering this contribution gives zero. This is easily understood,
since this term comes from the variation of $\chi$ in
$(\alpha g_{\mu\nu} - \gamma G_{\mu\nu}) \del^\mu \chi \del^\nu \chi $.
The contributions that survive are then given by
\bea
(\delta Q -i_\xi \Theta)_{\rm total} &=& -  r^{n-2} \fft{n-2}{r} \sqrt{\frac{h}{f}} \Big(\kappa \delta f  + \fft{3\gamma}{4} f \chi'^2 \delta f + \fft{\gamma}{2} f^2 \delta (\chi'^2) \Big ) \, \Omega_{(n-2)} \cr
& & -  r^{n-2}  \sqrt{\frac{h}{f}}  \kappa\,  \Big(\fft{f}{h}\,
\phi\delta\phi' + \fft{\phi \phi'}{2} (\fft{\delta f}{h} -
\fft{f \delta h}{h^2})\Big) \, \Omega_{(n-2)}\,.
\eea
For our solution, we can write $\delta (\chi'^2)$ in terms of $\delta f$,
and so
\bea
(\delta Q -i_\xi \Theta)_{\rm total} &=& -  r^{n-2} \fft{n-2}{r} \sqrt{\frac{h}{f}} \Big(\kappa  + \fft{\gamma}{4} f \chi'^2  \Big ) \delta f  \, \Omega_{(n-2)} \cr
& & -  r^{n-2}  \sqrt{\frac{h}{f}}  \kappa\,  \Big(\fft{f}{h}\,
\phi\delta\phi' + \fft{\phi \phi'}{2} (\fft{\delta f}{h} -
\fft{f \delta h}{h^2})\Big) \, \Omega_{(n-2)}\,.
\eea
It is now straightforward to verify that the Wald identity (\ref{waldid})
is indeed satisfied for all the charged black holes in section \ref{sec:d4bh}.

   We now turn to the question of interpreting  (\ref{waldid}) as the
first law of thermodynamics for these black hole solutions.
Let us first consider $\delta {\cal H}_\infty$.  We shall choose a gauge
for the Maxwell field such that it vanishes on the horizon.  We then find,
as we shall see later, that
\be
\delta \cH_\infty = \delta M - \Phi \delta Q \,.\label{deltaHinfinity}
\ee
where $\Phi$ is the electric potential at infinity.  More
precisely, $\Phi$ is the potential difference between the horizon and
asymptotic infinity, since $A$ vanishes on the horizon in our gauge choice.
The quantities $M$ and $Q$ can be interpreted as the mass and electric charge
of the solution.  This result is the same as in the case of
RN (AdS) black holes.

The situation is rather different from that for RN black holes when
we consider $\delta {\cal H}_+$ on the horizon.  This can be
studied in a rather general way, as follows.  We suppose that the
black hole event horizon is located at $r=r_0$, and that
the metric functions $h$ and $f$ near the horizon have Taylor expansions
of the form
\bea
f = f_1 (r - r_0) + f_2 (r-r_0)^2+\dots \,, \qquad
h =  h_1 (r - r_0) + h_2 (r-r_0)^2+\dots \,.
\eea
(It is easy to check that such expansions indeed arise for all the solutions
we have presented.)
For the class of black hole solutions we have considered, the scalar
field $\chi$ has the near-horizon behaviour
$\chi' \sim \fft {1}{ \sqrt {f } }$, and so in the near-horizon region
the function $\chi$ has an expansion of form
\be
\chi=\tilde \chi_0 + \tilde \chi_1 (r-r_0)^{\fft12} +
\tilde \chi_2 (r-r_0)^{\fft32} + \cdots\,.\label{phiexpan}
\ee
The electrostatic potential $\phi$, on the other hand, vanishes on the
horizon in our gauge choice.

   Taking the above expansions into account, we find that $\delta \cH$
on the horizon is given by
\be
\delta \cH_+ = \big(\kappa + \ft14 {\gamma}f \chi'^2\big|_{r_0}\big)\, T\,
\delta \big(\fft {\cal A} 4\big)\,,
\ee
where ${\cal A}=r_0^2\,\omega^{\phantom\Sigma}_{2,\epsilon}$
is the area of the horizon, and $T = \fft{\sqrt{f_1 h_1}}{4 \pi} $ is
the Hawking temperature.  This result was first obtained in
\cite{unchargedhorndeski} for neutral black holes in Einstein-Horndeski
gravity.  The formula remains unchanged for the charged black holes
we are considering here, provided we make our gauge choice in
which $\phi=0$ on the horizon.

   In the case of the neutral black holes considered in
\cite{unchargedhorndeski} , the solutions are characterised by a
single parameter, which may be taken to be the horizon radius $r_0$,
and so the expression
\be
\big(\kappa + \ft14 {\gamma}f \chi'^2\big|_{r_0}\big)\delta \big(\fft {\cal A} 4\big)=\delta S\,,
\ee
can always be integrated,
so that $\delta \cH_+=T\,\delta S$.  This was done in
\cite{unchargedhorndeski}, although questions about the validity of
interpreting the resulting function $S$ as the entropy were also
raised there, in view of the fact that there was no non-trivial
integrability check and furthermore that there was no independent
calculation of the entropy.

  For the charged black holes, we find that the differential
\be
\big(\kappa + \ft14 {\gamma}f \chi'^2\big|_{r_0}\big)\delta
\big(\fft {\cal A} 4\big)
\ee
cannot be interpreted as $\delta S$ for any function $S$, since it
is now a variation in a two-dimensional parameter space,
and this variation, thought of as a 1-form, is not an exact one.
One may take the two parameters to be the horizon radius $r_0$ and the
electric charge $Q_e$ of the black hole.
The non-integrability is a consequence of the fact that $f\chi'^2|_{r_0}$
is a function of both $r_0$ and $Q_e$, whilst the area of the horizon
depends only on $r_0$.

    To resolve this issue, we introduce the concept of a ``scalar charge''
defined by $Q_\chi=\int \sqrt{(\partial\chi)^2}$, where the integration is
over the level surfaces $d\Omega_{2,\epsilon}$.  This charge can be
evaluated both on the horizon and at infinity:
\bea
Q_\chi^+ &=&\int_{r=r_0} d\Omega_{2,\epsilon}\, \sqrt{(\partial\chi)_{+}^2}=
\omega_{2,\epsilon} \sqrt{f}\,\chi'\Big|_{r=r_0}\,,\cr
Q^\infty_{\chi} &=& \int_{r\rightarrow\infty} d\Omega_{2,\epsilon}\,
\sqrt{(\partial\chi)_{\infty}^2}=\omega_{2,\epsilon}\sqrt{f}\,\chi'\Big|_{r=\infty}\,.
\eea
An analogous kind of definition was used previously in the literature
in the case of a
(non-conserved) Yang-Mills ``charge''
\cite{Corichi:2000dm,Kleihaus:2002ee}, i.e.
\be
P^{\rm YM}=\int_{r\rightarrow \infty}
\sqrt{F^a_{\theta\phi} F^a_{\theta\phi}} \,d\theta d\varphi\,.
\ee
Note that this function of the Yang-Mills field vanishes when evaluated on
the horizon. It was shown \cite{Fan:2014ixa} that this definition of a
global Yang-Mills charge is indeed consistent with the first law of
thermodynamics, by using the Wald-formalism approach. The first law of AdS-Yang-Mills black holes were shown to be \cite{Fan:2014ixa}
\be
dM=TdS + \Phi_{\rm YM}\, dP^{\rm YM}\,.
\ee
In fact the electric charge can be defined in analogous way also, namely
\be
\fft{1}{\omega_{2,\epsilon}} \int \sqrt{|F|^2} d\Sigma =
\fft{1}{\omega_{2,\epsilon}} \int \sqrt{2}\, q d\Omega_{2,\epsilon}=\sqrt2\, q\propto Q_e\,.
\ee

   With this new scalar ``charge'' $Q_\chi^+$ on the horizon, we can write
$\delta {\cal H}_+$ in our case as
\be
\delta \cH_+ =
(\kappa + \fft {\gamma} {4} f \chi'^2|_{r_0}) T \delta (\fft {\cal A} 4) =
T \delta S +
\Phi_\chi^+ \delta Q_{\chi}^+\,,
\ee
where the entropy and the conjugate potential $\Phi_\chi$
for the scalar charge are given by
\be
S = \Big(\kappa + \fft {\gamma} {4} (f \chi'^2)|_{r_0} \Big)\fft {\cal A} 4 \,,\qquad
\Phi_\chi^+ = -\fft{\gamma\,{\cal A}\, T }{8\omega_{2,\epsilon}}\sqrt{f \chi'^2|_{r_0}}\,.\label{entropy}
\ee
The entropy has the standard contribution of one quarter of
the area of the horizon, plus a modification involving the scalar field
$\chi$.  Having interpreted both $\delta {\cal H}_\infty$ and
$\delta {\cal H}_+$, the Wald identity (\ref{waldid}) then yields the
first law in the form
\be
d M = T d S + \Phi_e d Q_e + \Phi_\chi^+ d Q^+_\chi\,. \label{1stlaw}
\ee
It is worth commenting that an analogous contribution from the scalar in the
last term of the above first law were also derived for general scalar
hairy AdS black holes \cite{Lu:2013ura,llscalar,Liu:2014tra}.  An important
difference is that in the previous examples, this extra term arose
in $\delta {\cal H}_\infty$, rather than $\delta {\cal H}_+$ as it does here.

To recapitulate, before we analyse the explicit examples, the Wald
identity (\ref{waldid}) is satisfied for the all black holes discussed in
section 3.  To interpret it as a first law, we find that the last
term in (\ref{1stlaw}) is unavoidable, since for the gauge choice where
the vector potential $dA$ vanishes on the horizon, the quantity
$T^{-1}\, \delta{\cal H}_+$ is not integrable.  In the first
law (\ref{1stlaw}), three quantities are unambiguous; they are the
Hawking temperature $T$, the electrostatic potential $\Phi_e$ and
the electric charge $Q_e$.  The other quantities are less clear cut, and
hence our
interpretation is not unique, but in any case the introduction of a
contribution from a global ``charge'' for the scalar $\chi$
is unavoidable.  We have also examined other possibilities, and
the first law as written in (\ref{1stlaw}) seems to give
the most reasonable interpretation of the Wald identity.  As we shall see
presently, for all the charged black holes the entropy defined as above
has the property that $S\sim T$ in the ``extremal'' limit
where $T\rightarrow 0$. Some further justifications will be
 given in sections 6 and 7.

We now turn to the explicit solutions.

\medskip
\noindent{\underline{\bf Asymptotically-Minkowski black holes }}
\medskip

For the asymptotically-Minkowski solutions (\ref{minso}), we
find that $\delta \cH$ evaluated at infinity and on the horizon are given by
\bea
\delta \cH_\infty = \ft12 {\kappa}  \delta \mu + \ft{1}4
\phi_0 \, \kappa \, \delta q\,,\qquad
\delta \cH_+ = \frac{\kappa  \left(q^2 - 8 {r_0}^2\right)^2
\delta r_0}{128  r_0^{4}}\,.
\eea
It is easy to verify that the Wald identity (\ref{waldid}) is indeed satisfied.  Following the earlier discussion, the thermodynamical quantities are given by
\bea
M &=& \ft12 \kappa \mu \,, \qquad \Phi_e =\fft{q(24r_0^2-q^2)}{24r_0^3} \,,
\qquad Q_e = \ft14 \kappa q \,,\cr
T &=& \frac{8 r_0^2 - q^2 }{32 \pi  r_0^3}\,, \qquad
S = \ft{1}{8} \pi \kappa  \left(8 r_0^2 - q^2\right)=4\kappa\, \pi^2 r_0^3\, T \,,\cr
Q_\chi^+ &=& 4\pi\sqrt{-\ft{\kappa  q^2}{2 \gamma r_0^2}} \,,\qquad
\Phi_\chi^+=-\ft18 \gamma r_0^2 T\, \sqrt{-\ft{\kappa  q^2}{2 \gamma r_0^2}} \,.
\eea
It easy to verify that the first law (\ref{1stlaw}) is satisfied. Note
that for the solution in this case to be real, we must require that
$\gamma<0$.  This implies that $\Phi_\chi^+$ is positive.

Note that the requirement that the curvature singularity at
$r_*\equiv q/\sqrt8$ lies inside the horizon (i.e. $r_*<r_0$) not only
guarantees that the temperature $T$ is positive, but also that the
entropy $S$ is positive.  The requirement also implies that
\be
M> M_*\equiv\ft{4\sqrt2}{3} Q_e\,.
\ee
Unlike the RN black hole, the mass/charge bound cannot be saturated.
As $M$ approaches $M_*$, both the temperature and entropy approach zero,
and $Q_\chi^+$ becomes a numerical constant.  The first law then
reduces to
\be
dM\rightarrow \Phi_e dQ_e\rightarrow dM_*\,,\qquad \hbox{with}\qquad \Phi_e\rightarrow \ft{4\sqrt2}{3}\,.
\ee

\bigskip
\noindent{\underline{\bf AdS planar black holes ($\epsilon = 0$)}}
\medskip

  Substituting the asymptotically-AdS planar black hole
solutions (\ref{ads0}) into the Wald formula gives
\bea
\delta \cH_\infty &=& \ft{1}{32 \pi} (4 \kappa + \beta  \gamma  ) \,
\delta \mu +\ft{1}{16 \pi}\phi_0 \,  \kappa \, \delta q\,,\quad
\delta \cH_+
= \frac{\left(\kappa  q^2-6 g^2 {r_0}^4 (\beta  \gamma +4 \kappa )
\right)^2 \delta {r_0} }{384 \pi  g^2 {r_0}^6 (\beta  \gamma +4 \kappa )}\,.
\eea
Note that we have chosen $\omega_{2,0}=1$ (a unit area 2-torus), so the
extensive quantities here may in general be interpreted as densities.
It is easy to verify that the Wald identity (\ref{waldid}) is satisfied,
from which we can read off the thermodynamical quantities
\bea
M &=& \ft{1}{32 \pi} (4 \kappa + \beta  \gamma  )  \mu \,, \qquad \Phi_e
= \phi_0 \,, \qquad Q_e = \ft{1}{16 \pi}\kappa q \,, \cr
T &=&\fft{3g^2 r_0}{4\pi} - \fft{\kappa\,q^2}{8(4\kappa + \beta\gamma) \pi r_0^3} \,, \qquad S =\fft{(4\kappa + \beta\gamma)\pi r_0}{12g^2} T\,,\nn\\
Q_\chi^+ &=& \sqrt{\beta -\frac{\kappa  q^2}{6 \gamma  g^2 r_0^4}}\,,\qquad
\Phi_\chi^+= -\ft18\gamma\,r_0^2\, T\, Q_\chi^+\,.
\eea
Here too, the requirement that the additional curvature singularity
at $r=r_*$ should lie inside the horizon ensures that both the temperature
$T$ and entropy $S$ are positive. This condition also implies that
\be
M>M_*\equiv \fft{32\times 2^{\fft14} \sqrt{\pi g}(4\kappa +
    \beta\gamma)^{\fft14}}{
5\times (3\kappa)^{\fft34}}\, Q^{\fft32}\,.
\ee
The equality cannot be saturated.  As $M$ approaches $M_*$, both the
temperature and entropy vanish, and furthermore $Q_\chi^+$ becomes a
pure numerical constant.  The first law then reduces to
\be
dM\rightarrow \Phi_e dQ_e \sim dM_*\,.\label{pseudoextremal}
\ee

\bigskip
\noindent{\underline{\bf (Lacally) AdS black holes ($\epsilon = 1$)} }
\medskip

For spherically-symmetric solutions, we have $\omega_{2,1}=4\pi$.  We find that the Wald formalism gives
\bea
\delta \cH_\infty &=& \ft{1}{8} (4 \kappa + \beta  \gamma)
\,\delta \mu -\ft{1}{4}\phi_0 \, \kappa \, \delta q\,,\cr
\delta \cH_+ &=& \frac{ \left(6 g^2 r_0^4 (4 \kappa  +\beta  \gamma)+ \kappa  \left(8 r_0^2-q^2\right)\right)^2 \delta r_0}{32
r_0^4 \left(3 g^2 r_0^2+1\right) (4 \kappa+\beta  \gamma )}\,,
\eea
and from these we can read off the thermodynamical quantities, namely
\bea
M &=& \ft{1}{8} (4 \kappa + \beta  \gamma  )  \mu \,,
 \qquad \Phi = \phi_0 \,, \qquad Q = \ft14 \kappa\, q \,,  \cr
T &=& \fft{3g^2 r_0}{4\pi} + \fft{\kappa (8r_0^2 - q^2)}{8(4\kappa + \beta
\gamma) \pi r_0^3}\,, \quad S = \fft{(4\kappa + \beta\gamma)\pi^2 r_0^3}{1+ 3g^2 r_0^2} T \,,\cr
Q_\chi^+ &=& 4\pi\sqrt{-\ft{\kappa q^2-6\beta\gamma g^2 r_0^4}{2\gamma r_0^2 (1 + 3g^2 r_0^2)}}\,,\qquad \Phi_\chi^+=-\fft{\gamma r_0^2}{32\pi} T\,Q_\chi^+\,.
\eea
There is a smooth limit under which $g \rightarrow 0$,
leading  back to the asymptotically flat case.  Again here, the requirement
that the singularity at $r=r_*$ lie inside the horizon ensures that both
the temperature and the entropy are positive.  It implies that $M>M_*$
where $M_*$ is a complicated function of the electric charge $Q_e$.
As $M\rightarrow M_*$, we have
$S\sim T\rightarrow 0$, so $Q_\chi^+$ becomes a purely numerical constant and
the first law reduces to (\ref{pseudoextremal}).

\section{Generalization to General Dimensions}

In the previous sections, we studied three classes of
static black hole solutions  in the four-dimensional charged Horndeski theory,
and obtained the first law of thermodynamics for
these black holes. We shall now generalize these results to
an arbitrary spacetime dimension $n$.

\bigskip
\noindent{\underline{\bf Asymptotically-Minkowski black holes }}
\medskip

The solution in a general dimension $n$ is given by
\bea
\chi'^2&=&-\frac{\kappa  q^2}{\gamma (n-2)(n-3)  r^{2 (n-3)} \, f}\,,
\qquad f = \frac{16 (n-2)^2 (n-3)^2 r^{4 n}}{\left(q^2 r^6-
  4 (n-2)(n-3) r^{2 n}\right)^2}\,h\,,\cr
h&=&1 - \fft{\mu}{r^{n-3}} +\frac{q^2}{2 (n-2)(n-3))r^{2 (n-3)} }-
\frac{q^4}{48 (n-2)^2 (n-3)^2 r^{4 (n-3)}}\,,\cr
\phi &=&\phi_0 - \frac{q}{(n-3) r^{n-3}}-
  \frac{q^3}{12 (n-2)(n-3)^2 r^{3 (n-3)}}\,.
\eea
The solutions contain two non-trivial parameters, which we
make take to be $\mu$ and $q$.
As we shall see presently, these parameterise the mass and the
electric charge respectively.  The further integration constant
$\phi_0$ is a pure
gauge parameter, and as before we shall adopt the gauge choice
where the potential $A$ vanishes on the horizon.  The solution contains
two curvature singularities: the usual one at the origin $r=0$, and the
other located at $r=r_*$, given by
\be
4 (n-2)(n-3) r_*^{2 n - 6}-q^2 = 0 \,.
\ee
The event horizon $r=r_0$ is at the largest root of $h=0$.  Furthermore,
we must have $r_0>r_*$, which implies
\be
\fft{\mu}{q}> \fft{4}{3\sqrt{(n-2)(n-3)}}\,.
\ee
Applying the Wald formalism, we find
\bea
\delta \cH_\infty &=& \frac{\kappa}{16 \pi} (n-2) \,\delta \mu
  -\fft {1}{16 \pi}\phi_0 \, \kappa \, \delta q\,,\cr
\delta \cH_+ &=& \frac{\kappa \left(q^2 - 4 (n-2)(n-3)
{r_0}^{2 n - 6}\right)^2 \delta r_0 }{16 (n-2) (n-3)
{r_0}^{3 n - 8} }\,.
\eea
It is then easy to establish that the Wald identity (\ref{waldid}) is satisfied. This leads to the first law (\ref{1stlaw}) with
\bea
M &=& \frac{\kappa}{16 \pi} (n-2) \omega\,\mu \,, \quad \Phi_e =\fft{q\big( 12(n-2)(n-3) r_0^{2(n-3)} - q^2\big)}{12(n-2)(n-3)^2\, r_0^{3(n-3)}}\,, \quad Q_e = \fft{\kappa}{16 \pi} \omega\,q \,, \cr
T &=& \frac{4 (n-2)(n-3)r_0^{2 n - 6} - q^2 }{16 \pi  (n-2)
r_0^{2 n - 5}  } \,,\quad
S = \fft{\kappa \pi \omega}{n-3}\, r_0^{n-1} T\,,\cr
Q_\chi^+ &=& \sqrt{-\frac{\kappa  q^2 }{\gamma
(n-2)(n-3) {r_0}^{2 n - 6} }}\,, \qquad \Phi_\chi^+=-\fft{\gamma}{8\omega} r_0^{n-2} \,T\,Q_\chi^+\,,
\eea
where $\omega$ is the volume of the unit $S^{n-2}$.

\bigskip
\noindent{\underline{\bf Asymptotically-AdS planar solutions ($\epsilon = 0$)}}
\medskip

The solutions are given by
\bea
\phi&=&\phi_0-\frac{q}{(n-3) r^{n-3}}+\frac{\kappa  q^3}{
g^2 (n-1)(n-2)(3 n-7)(4 \kappa+\beta  \gamma )r^{3 n-7}}\,,\cr
\chi' &=& \sqrt{\beta -\frac{\kappa  q^2}{\gamma
g^2 (n-1)(n-2) r^{2 (n-2)}}}\,\fft{1}{\sqrt{f}}\,,\qquad  \cr
f&=& \frac{g^4 (n-1)^2 (n-2)^2 r^{4 n} (4 \kappa+\beta\gamma )^2}{
\left(\kappa  q^2 r^4-g^2 (n-1)(n-2) r^{2 n} (4\kappa+\beta\gamma )\right)^2} \, h\,,\cr
h&=& g^2 r^2 - \fft{\mu}{r^{n-3}} +\frac{2 \kappa  q^2}{(n-2) (n-3)
(4\kappa+\beta\gamma)  r^{2 (n-3)}}  \cr
&&-\frac{\kappa ^2 q^4}{g^2 (n-1)(3 n -7) (n-2)^2(4\kappa+\beta\gamma)^2 r^{4 n -10}}\,. \label{plaads}
\eea
Besides the usual curvature singularity $r=0$, there is also an
additional curvature singularity $r_*$, which is given by
\be
(n-1)(n-2) g^2 r_*^{2 n-4} (\beta  \gamma +4 \kappa )-\kappa  q^2 = 0 \,.
\ee
For the solution to be free from naked curvature singularities,
namely the event horizon must be outside the second singularity,
$r=r_0> r_*$, we must have
\be
\mu > \fft{2^{\fft{2n-5}{n-2}} (n-2)^{\fft{3n-7}{2(n-2)}} g^{\fft{n-3}{n-2}}}{(n-3)(3n-7)(n-1)^{\fft{n-1}{2(n-2)}}}\Big(1 + \fft{\beta\gamma}{4\kappa}\Big)^{-\fft{n-1}{2(n-2)}} q^{\fft{n-1}{n-2}}\,.
\ee
Evaluating $\delta \cH$ both at infinity and on the horizon gives
\bea
\delta \cH_\infty &=& \frac{1}{64 \pi} (n-2) (\beta  \gamma +4 \kappa )
\,\delta \mu-\fft {1}{16 \pi}\phi_0 \,\kappa  \,\delta q\,,\cr
\delta \cH_+ &=& \frac{  \left(\kappa  q^2  -
(n-1)(n-2) g^2 {r_0}^{2 n - 4} (\beta  \gamma +4 \kappa )\right)^2
\delta r_0}{64 \pi g^2 (n-1) (n-2) (\beta  \gamma +4 \kappa ){r_0}^{3 n- 6}}\,.
\eea
We find that indeed the Wald identity (\ref{waldid}) is satisfied.
The first law of thermodynamics is then given by (\ref{1stlaw}), with
\bea
M &=& \frac{1}{64 \pi} (n-2) (\beta  \gamma +4 \kappa )  \mu \,, \qquad
Q_e = \fft{\kappa}{16 \pi} q \,,\cr
\Phi_e &=&
\fft{q r_0^{3-n}}{n-3} -\fft{\kappa q^3 r_0^{7-3n}}{(n-1)(n-2) (3n-7) g^2 (4\kappa + \beta \gamma)}\,,\cr
T &=& \fft{(n-1)g^2 r_0}{4\pi} - \fft{\kappa q^2 r_0^{5-2n}}{4(n-2)\pi (4\kappa +\beta\gamma)}\,,\qquad S = \fft{(4\kappa + \beta\gamma) \pi}{4(n-1) g^2}\,
r_0^{n-3}\,T\,,\cr
Q_\chi^+ &=& \sqrt{\beta - \fft{\kappa q^2 r_0^{2(2-n)}}{(n-1)(n-2) g^2 \gamma}}\,,\qquad \Phi_\chi^+=-\fft{\gamma}{8} r_0^{n-2} T Q_\chi^+\,.
\eea
Note that we have taken the $(n-2)$-torus sections to have unit area,
and so in general the extensive quantities $(M, S, Q_e, Q_\chi^+)$ are
to be interpreted as densities.

\bigskip
\noindent{\underline{\bf Asymptotically-AdS spherical
 black holes ($\epsilon = 1$)} }
\medskip

With the spherically-symmetric ansatz, we may follow the same procedure
to solve the equations of motion and obtain the solution in a general
dimension $n$. The solutions now are more complicated and for a general
dimensions $n$ they
are naturally expressed in terms of hypergeometric functions.
The $f$ and  $\chi$ functions are related to $h$ as usual, and are
given by
 \bea
 \fft{f}{h} &=& \frac{(n-2)^2(4\kappa + \beta\gamma)^2 \big((n-1)g^2r^2+n-3\big)^2}{\big((n-1)(n-2) (4\kappa + \beta\gamma)g^2r^{2}+\kappa  \left(4 (n-2)(n-3)-q^2 r^{2(3-n)}\right)\big)^2}\,,\cr
 \chi' &=&  \sqrt { \frac{ \beta  \gamma  g^2(n-1)(n-2) r^{2 n - 4}-
 \kappa q^2 }{\gamma  (n-2) \big((n-1)g^2r^2+n-3\big) r^{2 (n - 3)}}  }
 \sqrt{\fft1 f}\,.
 \eea
Once $f/h$ is known, the function $\phi$ can be easily integrated out,
and is given by
\bea
\phi  &=& \phi_0 -\frac{q }{(n-3) r^{n - 3}} +\frac{\beta  \gamma  (n-3) q }{g^2 (n-1)^2 (4\kappa+\beta\gamma) r^{n - 1}}\cr
&&-\frac{\beta  \gamma  (n-3)^2 q  \, _2F_1\left(1,\frac{n+1}{2};\frac{n+3}{2};\frac{3-n}{(n-1) g^2r^2}\right)}{g^4 (n-1)^2 (n+1) (4\kappa+\beta\gamma) r^{n+1} } \cr
 && + \frac{\kappa  q^3  \, _2F_1\left(1,\frac{1}{2} (3 n-7);
\frac{1}{2} (3 n-5);\frac{3-n}{(n-1) g^2 r^2}\right)}{g^2
(n-1) (n-2) (3 n-7) (4\kappa+\beta\gamma) r^{3 n - 7}}  \,.
\eea
The integration constant $\phi_0$ is pure gauge and we shall as usual
choose it so that $\phi$ vanishes on the horizon. The function $h$ is
given by $h=\bar h + h_q$ with
\bea
h_q &=&\frac{2 \kappa  q^2 }{(n-2) (n-3) (4\kappa+\beta\gamma) r^{2 n - 6}}  -\frac{2 \beta  \gamma  \kappa  (n-3) q^2 }{g^2 (n-1)^2 (n-2) (4\kappa+\beta\gamma)^2 r^{2 n - 4} } \cr
&& +\frac{2 \beta  \gamma  \kappa  (n-3)^2 q^2 \, _2F_1\left(1,\frac{n+1}{2};\frac{n+3}{2};\frac{3-n}{(n-1) g^2r^2}\right)}{g^4 (n+1)(n-1)^2 (n-2)(4\kappa+\beta\gamma)^2 r^{2 n -2 } }  \cr
&& - \frac{\kappa ^2 q^4  \, _2F_1\left(1,\frac{1}{2} (3 n-7);\frac{1}{2} (3 n-5);\frac{3-n}{(n-1) g^2r^2}\right)}{g^2 (n-1)(n-2)^2(3 n-7) (4\kappa+\beta\gamma)^2 r^{2(2n -5)}}\,.\label{hqexp}
\eea

   The hypergeometric functions in (\ref{hqexp})
are well defined for all integers $n\ge 4$.  For $\bar h$, it is more
convenient to present separate expressions in even and in odd dimensions.
When $n$ is even, we find
\bea
\bar h &=&
\fft{8g^2\kappa (2\kappa + \beta\gamma)}{(4\kappa + \beta\gamma)^2} r^2 +
\fft{16\kappa^2}{(4\kappa + \beta\gamma)^2} - \fft{\mu}{r^{n-3}}\cr
&&+\fft{\beta^2\gamma^2 g^2r^2}{(4\kappa + \beta\gamma)^2}\, {}_2F_1[
1,-\ft12(n-1); -\ft12(n-3); \ft{3-n}{(n-1)g^2r^2}]\,.
\eea
This expression becomes divergent if $n$ is an odd integer.  The divergence
occurs because the third argument in each of
these two functions becomes a negative integer.  This can be handled by
noting that the hypergeometric
functions are of the special form $_2F_1(1,b; b+1; -x)$, and making
use of the identity
\be
_2F_1[1,b;b+1; -x] = \fft{b}{(b-1)x}\, _2F_1[1,1-b; 2-b; -\fft1{x}]
 +\pi b\csc\pi b\, x^{-b}\,,\label{hype}
\ee
in the term where the problem arises.  It is then evident that the
divergence comes purely from the $\pi b\csc\pi b\, x^{-b}$ term, and
furthermore, that these lead to terms only at order
$1/r^{n-3}$ in
the expression for $h$, with diverging coefficients. The infinities
can therefore be absorbed into an infinite additive renormalisation
of the
mass coefficient $\mu$ in the $-\mu/r^{n-3}$ term in $h$.  By using the
identity (\ref{hype}) and redefining the integration constants, we get
\bea
\bar h &=& \fft{8g^2\kappa (2\kappa + \beta\gamma)}{(4\kappa
                              + \beta\gamma)^2} r^2 +
\fft{16\kappa^2}{(4\kappa + \beta\gamma)^2} - \fft{\mu}{r^{n-3}}\cr
&&+\fft{(n-1)^2 \beta^2\gamma^2 g^4r^4}{(n-1)(n-3)(4\kappa + \beta\gamma)^2}\, {}_2F_1[
1,\ft12(n+1); \ft12(n+3); \ft{(n-1)g^2r^2}{3-n}]\,,
\eea
for odd $n\ge 5$.

Applying the Wald formalism, we find
\crampest{
\bea
\delta \cH_\infty &=& \frac{1}{64 \pi} (n-2) (\beta  \gamma +4 \kappa )
\,\delta \mu  -\fft {1}{16 \pi}\phi_0 \,\kappa  \,\delta q\,,\cr
\delta \cH_+ &=& \frac{ \left[(n-1)(n-2) g^2  {r_0}^{2 n+2}
(\beta  \gamma +4 \kappa )+\kappa  \left(4 (n-2)(n-3){r_0}^{2 n}
  -q^2 {r_0}^6\right)\right]^2 \delta r_0}{64 \pi  (n-2) (\beta  \gamma +
  4 \kappa ) \left[ (n-1)g^2  {r_0}^2 +n-3\right]
{r_0}^{3 n + 4}}\,.
\eea}
It is then straightforward to verify that the Wald identity (\ref{waldid})
is indeed satisfied by the general spherically-symmetric black holes.
With this identity we can establish the first law (\ref{1stlaw}), with
\bea
M &=& \frac{\omega}{64 \pi} (n-2) (\beta  \gamma +4 \kappa )  \mu \,,
\quad \Phi = \phi_0 \,, \quad Q = \fft{\kappa\omega}{16 \pi} q \,, \cr
T &=& \fft{(n-1)g^2r_0}{4\pi} +
\fft{\kappa\big(4(n-2)(n-3) r_0^{2(n-3)} - q^2\big)}{4\pi(n-2) (4\kappa + \beta
\gamma)r_0^{2n-5}}\,,\cr
S &=& \fft{\pi(4\kappa + \beta\gamma)\omega}{4\big( (n-1)g^2r_0^2 + n-3\big)}
r_0^{n-1} T\,,\cr
Q_\chi^+ &=&\omega \sqrt{-\ft{\kappa q^2 - (n-1)(n-2)\beta\gamma g^2 r_0^{2(n-2)}}{
(n-2)\gamma ((n-1) g^2 r_0^2 +n-3) r_0^{2(n-3)}}}\,,\qquad
\Phi_\chi^+ = -\fft{\gamma}{8\omega} r_0^{n-2} T Q_\chi^+\,.
\eea
Here $\omega$ is the volume of the unit $S^{n-2}$.  It is worth noting
that the general forms of the thermodynamical quantities are the same
for both even and odd dimensions.

The following properties apply to all the solutions discussed above.  In
addition to $r=0$, a spacetime curvature singularity can also arise at
$r_*$, which is the root of $h/f$.  The requirement that $r_*<r_0$ implies
that $M>M_*$ where $M_*$ is some function of the electric charge
$Q_e$. As $M\rightarrow M_*$, we have
$S\sim T\rightarrow 0$.  The scalar charge $Q_\chi^+$ can be expressed as
\be
(Q_\chi^+)^2=-\fft{4\kappa}{\gamma} + \fft{16}{\gamma\, {\cal A}} S\,,
\ee
which becomes purely numerical as $S\rightarrow 0$.  Thus as
$M\rightarrow M_*$, the first law reduces to (\ref{pseudoextremal}).

\section{Noether Charge and Scaling Symmetry}

Having obtained the first law of thermodynamics (\ref{1stlaw}) from the
Wald identity (\ref{waldid}), we shall probe the results further in this
and the next section.  As we have seen, the right-hand side of the
first law involves six
thermodynamical quantities, of which three can be unambiguously
defined independently
of the Wald formalism. These are the Hawking temperature $T$,
and the electrostatic potential and charge $(\Phi_e,Q_e)$.  Our interpretation
for the definitions of the other three quantities; the entropy $S$ and
$(\Phi_\chi^+, Q_\chi^+)$, and also the mass $M$,  would be
more satisfactory if we could find a different method to calculate
them.  In this section, we shall examine these quantities from a
different angle.

    As discussed in \cite{Liu:2015tqa} and also in \cite{unchargedhorndeski},
the existence of a scaling symmetry for AdS planar black holes has the
important consequence that there exists a generalized Smarr formula,
which relates the thermodynamic mass $M$ to the other thermodynamical
quantities. To see the
scaling property more explicitly, we rewrite the ansatz for the AdS planar
black-hole solutions in the form
\be
ds^2 = d\rho^2 - a(\rho)^2dt^2 + b(\rho)^2 (dx^i dx^i)\,,\qquad
\chi=\chi(\rho)\,, \qquad \phi = \phi(\rho)\,,
\ee
where the new radial coordinate $\rho$ is related to $r$ by
$f^{-1/2}\, dr = d\rho$.
The equations of motion can be derived from
the effective one-dimensional Lagrangian
\bea
L&=& \fft{1}{16\pi} ab^{n-2}\Big[\kappa (R + \fft {\dot\phi^2}{2 a^2}
- 2\Lambda)
- \ft12\alpha \dot \chi^2 + \ft12
\gamma G_{11} \dot \chi^2\Big]\,,\cr
R &=& -\fft{2\ddot a}{a} - \fft{2(n-2)\ddot b}{b} -
\fft{2(n-2)\dot a\dot b}{ab} - \fft{(n-2)(n-3) \dot b^2}{b^2} \,,\cr
G_{11} &=& \fft{(n-2)\dot a \dot b}{ab} + \fft{(n-2)(n-3) \dot b^2}{2 b^2}\,,
\eea
where a dot denotes a derivative with respect to $\rho$.
The Lagrangian is invariant under the global scaling
\be
a\rightarrow \lambda^{2-n}\,a\,,\qquad b\rightarrow \lambda\, b \,, \qquad \phi \rightarrow \lambda^{2-n}\, \phi \,.
\ee

    This global symmetry implies that there is an associated
conserved Noether charge, given by
\be
{\cal Q}_N=
\fft{1}{32\pi} (n-2) a b^{n-2} \Big[ (4\kappa + \gamma \dot\chi^2)
 (\fft {\dot a}{a}
  - \fft {\dot b}{b})  - \fft{2 \kappa \phi \dot \phi}{a^2}\Big]\,.
\ee
In terms of the coordinates of the original ansatz (\ref{staans}) with $\epsilon = 0$,
we have
\be
{\cal Q}_N
= \fft{1}{16 \pi} (n-2) r^{n-2} \Big(  \sqrt{hf} (\kappa + \fft{\gamma}{4 } f \chi'^2) (\fft{h'}{h} - \fft2r)\Big) - \kappa \sqrt{\fft fh} \phi \phi' \Big)\,.
\ee
Substituting the AdS planar black hole solution into this Noether charge
formula and evaluating at asymptotic infinity, we find
\be
{\cal Q}_N\big|_\infty = (n-1) M - (n-2) \Phi_e Q_e\,.\label{QNinf}
\ee
(Here we have, as usual, chosen the gauge where $A$ vanishes on the horizon.)
In this equation, $\Phi_e$ and $Q_e$ are defined unambiguously,
using the standard properties of the Maxwell field.  The mass $M$
read off from (\ref{QNinf}) turns out to be identical to the one we
obtained previously from the Wald formalism.  This result is the same
as that for the RN-AdS planar black hole, and it
can be viewed as an independent derivation of the mass.  (See also
\cite{peng} for another discussion of the calculation of the mass of
black holes in Horndeski gravity based on Wald formalism.)

If we evaluate the Noether charge on the horizon, we find
\be
{\cal Q}_N\big|_+  =(n-2)\, \ft{\sqrt{h'(r_0)f'(r_0)}}{4\pi}\, \Big(\kappa + \ft14\gamma (f \chi'^2)|_{r_0}\Big)\fft{\cal A}{4}=(n-2) T S\,,
\ee
where $S$ is precisely the entropy defined according to (\ref{entropy}),
thus strongly supporting the naturalness of this interpretation.
We therefore have the generalized Smarr formula
\be
 M =\fft{n-2}{n-1} \,  \Big( \Phi Q + TS \Big)\,, \label{sma}
\ee
which takes the same form as for the RN-AdS planar black holes
of Einstein-Maxwell theory. Note that the newly introduced
thermodynamical quantities $\Phi_\chi^+$ and $Q_\chi^+$
do not enter the generalized Smarr formula.  This is completely consistent
with their specific scaling behaviours.  Interestingly, it was shown that the scalar contributions via $\delta{\cal H}_\infty$ to the first law of the scalar hairy AdS-planar black holes do not enter the generalized Smarr relation either \cite{Liu:2015tqa}.

It is worth pointing out that in the neutral AdS planar black hole in
Einstein-Horndeski gravity, the quantity $f\chi'^2=\beta$ is a non-thermodynamical constant, and hence
the term $\Phi_\chi^+d Q_\chi^+$ does not enter the first law.  The necessity
of introducing a scalar charge $Q_\chi^+$ does not arise in that case.
Correspondingly, the first law can be simply stated as $dM=TdS$ with
the generalized Smarr relation $M=\fft{n-2}{n-1} \, TS$
\cite{unchargedhorndeski}.  However, when the electric charge is turned on,
the quantity $f\chi'^2$ ceases to be constant, and hence the
first law acquires the extra contribution, as in (\ref{1stlaw}).

\section{Viscosity/Entropy Ratio}

Based on the AdS/CFT correspondence, the AdS planar black holes can be
viewed as dual to some ideal fluid in the boundary theory. One can study
the linear response of the graviton modes and determine the shear
viscosity $\eta$ of the boundary fluid \cite{KSS,KSS0}.  One can then
calculate the viscosity/entropy ratio.  In typical situations, the
evaluation of the entropy is straightforward, and
it can be obtained simply by using the standard
Wald entropy formula (\ref{waldentropy}).  However, as we have seen,
in the EHM theory the determination of the entropy is far from trivial. In
this section, we shall first evaluate the viscosity using a standard
method in the literature, and then compute the $\eta/S$ ratio.  The elegance
of the result may be viewed as providing further support
for the naturalness of our entropy formula (\ref{entropy}).

To calculate the shear viscosity of the boundary field theory, we consider
a transverse and traceless perturbation of the AdS planar black hole, namely
\be
ds^2 =-h dt^2 + \fft{dr^2}{f} + r^2 \big(dx_i dx_i +
  2 \Psi(r,t) \, dx_1 dx_2\big)\,,
\ee
where the background solution is given by
(\ref{plaads}).  We find that the
mode $\Psi(r,t)$ satisfies the linearised equation
\bea
&&2 r h f \big( \kappa q^2 - (n-1)(n-2) (4 \kappa + \beta \gamma) g^2 r^{2 (n-2)} \big)\, \Psi'' \cr
 && +  2 r \big( (n-1)(n-2) (4 \kappa - \beta \gamma) g^2 r^{2 (n-2)}  + \kappa q^2 \big)\, \ddot \Psi\cr
&& - h \Big[  \, \, \Big( (3 n -7) \kappa q^2 + (n-1)^2(n-2) (4 \kappa + \beta \gamma ) g^2 r^{2 (n-2)}\Big) f  \cr
&& \qquad + r \Big( (n-1)(n-2) (4 \kappa + \beta \gamma) g^2 r^{2 (n-2)}
- \kappa q^2  \Big) \big ( (n-1) g^2 r + f'  \big)  \Big]\, \Psi' =0\,,
\label{perteom}
\eea
where we are using a prime to denote a derivative with respect to $r$,
and in this section a dot denotes a derivative with respect to $t$.
For an infalling wave which is purely ingoing at the horizon, the solution
for a wave with low frequency $\omega$ is given by
\bea
\Psi(r,t) &=& e^{-{\rm i} \omega t} \, \psi(r)\,,\qquad \psi(r)=
\exp\Big[-{\rm i} \omega K \log\fft{h(r)}{g^2 r^2}\Big] \,
 \big( 1 - i \omega U(r) + {\cal O}(\omega^2) \big)\,,\cr
U(r) &=&  1 -\int dr \, \fft{4 \kappa q K \phi}{ (4 \kappa + \beta \gamma )
r^{n-2} h} \,, \cr
\cr
K &=&\fft{1}{4\pi T} \sqrt{\fft{ (n-1)(n-2) (4 \kappa - \beta \gamma ) g^2 r_0^{2 (n-2)} + \kappa q^2}{ (n-1)(n-2) (4 \kappa + \beta \gamma ) g^2 r_0^{2 (n-2)} - \kappa q^2}}  \,. \label{pereq}
\eea
Note that the constant parameter $K$ is determined by the horizon
boundary condition.
The overall integration constant is fixed so that $\Psi$ is unimodular
asymptotically, as $r\rightarrow \infty$.

In order to study the boundary field theory using the AdS/CFT correspondence,
we substitute the ansatz with the linearised perturbation into the action.
The quadratic terms in the Lagrangian, after removing the
second-derivative contributions by including a Gibbons-Hawking term, can
be written as
\be
{\cal L}_2 = P_1\, {\Psi'}^2 + P_2\, \Psi\,\Psi' + P_3\, \Psi^2
+ P_4\, \dot\Psi^2 \,,
\ee
with
\bea
P_1 &=& - \fft18 r^{n-2} \sqrt{hf} \big( 4 \kappa + \gamma f \chi'^2 \big)\,, \quad P_2 =\fft14 r^{n-3} \sqrt{hf} \big[ - 8 \kappa + \gamma \big( (n-1) g^2 r^2 + (n-3) f \big) \chi'^2 \big] \,,  \cr
P_3 &=&  \fft{1}{4 r^{n-2}} \sqrt{\fft hf} \Big[ \,\, (n-1) (n-3) \kappa f r^{2 (n-3)} + (n-1) \kappa g^2 r^{2n - 3} \big( \fft{(n-1) g^2 r}{f} - \fft{f'}{f} \big)  \cr
&& \qquad \qquad \quad - \Big( \kappa q^2 + (n-1) g^2 r^{2 (n-2)} \big( 2 (n-4) \kappa + (n-2) \beta \gamma \big)  - (n-1) \kappa r^{2 n-5} f' \Big) \Big]\,, \cr
P_4 &=& \fft{3 r^{n-2}}{8 \sqrt{hf}} (- 4 \kappa + \gamma f\chi'^2 )
\eea
Note that $P_3 = \ft12 P_2'$.  We then find that the terms quadratic in
$\Psi$ in
the Lagrangian are given by
\be
{\cal L}_2 = \fft{d}{dr}\, (P_1 \Psi\,\Psi' + \ft12 P_2\Psi^2)
  +\fft{d}{dt} (P_4\, \Psi\, \dot\Psi) -
   \Psi\, \Big[ P_1\, \Psi'' + P_1'\, \Psi' + P_4\, \ddot{\Psi} \Big]\,.
\ee
The last term , enclosed in square brackets, vanishes by virtue of
the linearised perturbation equation (\ref{perteom}), and so the
quadratic Lagrangian is a total derivative.  The viscosity is determined
from the $P_1 \Psi \Psi'$ term,
following the procedure described in \cite{sonsta,shenker}.  Using
this, we find that the viscosity is given by
\be
\eta= \fft{K}{16 \pi } \Big[ \fft{n-1}{4} (4 \kappa + \beta \gamma ) \mu
- \kappa \phi_0 q \Big] = K \,\Big[ \fft{n-1}{n-2}M - \Phi Q \Big]\,.
\label{eta0}
\ee
According to the generalized Smarr formula (\ref{sma}), this implies
\be
\eta = K T S\,.
\ee
We therefore find that the viscosity/entropy ratio  is given by
\be
\fft{\eta}{S} = K T =  \fft{1}{4\pi}\sqrt{\fft{ (n-1)(n-2) (4 \kappa - \beta \gamma ) g^2 r_0^{2 (n-2)} + \kappa q^2}{ (n-1)(n-2) (4 \kappa + \beta \gamma ) g^2 r_0^{2 (n-2)} - \kappa q^2}} \label{eta1}
\ee
for the charged AdS planar Horndeski black holes.  Note that when $q=0$,
the result reduces to that for the neutral black hole,
which was obtained in \cite{unchargedhorndeski}. The viscosity/entropy
ratio can also be expressed as
\be
\fft{\eta}{S} = \fft{1}{4\pi} \sqrt{\fft{\kappa r_0^{n-2}}{2S}-1}
= \fft{1}{4\pi} \sqrt{\fft{2(n-1) \kappa g^2 r_0}{(4\kappa + \beta\gamma)\pi T}-1}\,.\label{eta2}
\ee
Although the final result (\ref{eta1}) looks quite complicated, it
still has a rather elegant form, especially when it is expressed in terms
of the entropy or temperature, as in (\ref{eta2}).

   The viscosity (\ref{eta0}) is given in terms of quantities
that are defined asymptotically at infinity, whose interpretations
are unambiguous.  The only thing that was uncertain previously
was the definition of the entropy $S$. Our formula (\ref{entropy}) provides
a simpler and more elegant form for the ratio (\ref{eta1})
than alternative possibilities. Nevertheless, the viscosity/entropy
ratio (\ref{eta1}) is still rather complicated.  It depends not only
on the parameters in the theory, such as $(\kappa, \beta,\gamma)$, but
also on the integration constants in the solution, namely the mass and
charge.  The ratio is dependent on $r_0$ and $T$, which is qualitatively
analogous to what was found for charged black holes in
Einstein-Gauss-Bonnet-Maxwell
theory. (See e.g.~\cite{Cai:2008ph,Hu:2011ze}.)  However, the detailed
dependence is quite different. For fixed charge $Q_e$, in the ``extremal''
limit as $T\rightarrow 0$, the viscosity and the entropy both vanish, as
\be
\eta\Big|_{T\rightarrow 0} \rightarrow \sqrt{\ft{\kappa (4\kappa +\beta\gamma) r_0^{2n-5}}{128 (n-1)\pi g^2}\, T}\,,\qquad
S\Big|_{T\rightarrow 0} \rightarrow \ft{(4\kappa + \beta\gamma) \pi r_0^{n-3}}{
4(n-1) g^2}\, T\,.
\ee
It follows that the viscosity/entropy ratio diverges in the
low temperature $T\rightarrow 0$ limit.  On the other hand, in the
large temperature limit $T\rightarrow \infty$, the $q$ terms in
(\ref{eta1}) become negligible, and $\eta/S$ approaches a constant
that is independent of the temperature, and in fact is equal to the
ratio obtained for the neutral black hole.

\section{Conclusions}

In this paper, we considered the Einstein-Horndeski-Maxwell (EHM)
theory where the derivative of
the axionic scalar $\chi$ couples non-minimally to the Einstein tensor,
as in (\ref{action}).  We studied the thermodynamics of the static black
hole solutions in this theory by using the Wald formalism, and we
found that the scalar field has a direct contribution to the first law.
The contribution of the scalar can be attributed  to a branch-cut singularity
in the behavior of the scalar on the horizon. However, the scalar is
axionic and enters the theory only through its derivative, and in an
orthonormal frame $\partial_a \chi$ is regular everywhere, both on and
outside the horizon, and all invariants involving the scalar field
are finite everywhere.

  In the Wald formalism, the first law of black hole thermodynamics can be
read off from the Wald identity (\ref{waldid}), provided one
can attach an appropriate thermodynamic meaning to the various
quantities appearing on the two sides of the equation.  We found that the
Wald identity is valid for all the charged solutions, supporting their
interpretation as regular
black holes.  However, it is a non-trivial matter to translate the Wald
identity into a first law, since to do this one needs to identify the
quantities appearing in the two sides of (\ref{waldid}) in terms of
thermodynamic variables. In a previous paper \cite{unchargedhorndeski}
where neutral solutions were studied, the static black holes had only one
non-trivial parameter, and so the integrability of $\delta \cH_+/T$  was
always guaranteed, allowing one to define an entropy and hence
obtain the first law $dM=TdS$ for these black holes.   For the charged
solutions considered in this paper, on
the other hand, there are two non-trivial parameters, namely the mass and
the charge, and we found that $\delta \cH_+/T$ was not an exact differential.
It therefore appears to
be unavoidable that one must introduce an additional pair of
thermodynamical quantities.  Inspired by the definition of a global
``charge'' in
Yang-Mills theory, we defined an analogous scalar ``charge'' $Q_\chi^+$
evaluated on the horizon, and also its thermodynamical conjugate potential
$\Phi_\chi^+$.  We were then able to obtain an extended first law of
thermodynamics (\ref{1stlaw}).

   We explicitly analysed  three kinds of static black-hole solutions in
four  dimensions, applied the Wald formalism, and obtained
the thermodynamic variables and derived the first law.
The black hole entropy is the usual one quarter of the horizon area
(which is the result that follows form the standard Wald entropy formula)
plus a modification from the scalar field. Apart from the usual curvature
singularity at $r=0$, a second curvature singularity at some radius
$r=r_*$ emerges in these black
hole solutions. For suitable choices of the parameter ranges the
second singularity lies inside the event horizon.  This also ensures
 that the Hawking temperature and the entropy are positive.  We also
obtained generalisations of all these results to arbitrary spacetime
dimensions.

We then tested the validity of our interpretation of the entropy
by considering independent methods for calculating the thermodynamical
quantities.  One such method, applicable to the case of planar black holes,
involved deriving the Noether charge associated with the scaling symmetry
of the planar black hole solutions.   We also calculated  the
viscosity/entropy ratio in the dual boundary theory for the planar
black hole examples, and
the generalized Smarr relation. These results provide some support for
the validity of our calculations of the  thermodynamical quantities for
the charged black holes.

There are several important differences between the charged black holes in
EHM theory and the RN black holes of Einstein-Maxwell theory.  Although
both have ``extremal'' limits where the temperature approaches zero,
the temperature of the charged black holes in the EHM theory can never
reach absolute zero.  Furthermore, as $T\rightarrow 0$, the black hole
entropy in EHM theory also approaches zero, rather than some non-vanishing
fixed value as in the case of RN black holes.  Thus the charged black
holes in EHM theory appear to have properties more in line
with conventional systems as far as the third law of thermodynamics
is concerned. Such characteristics of the black hole solutions
in the EHM theory have not previously been seen in the black holes
of more conventional theories, and hence it is of great interest to
investigate them further.

\section*{Acknowledgements}

We are grateful to Sera Cremonini for helpful discussions.
H-S.L.~is supported in part by NSFC grants 11305140, 11375153 and 11475148,
SFZJED grant Y201329687 and CSC scholarship No. 201408330017. C.N.P.~is supported in part by DOE grant DE-FG02-13ER42020. The work of X-H.Feng and H.L.~is supported in part by NSFC grants NO. 11175269, NO. 11475024 and NO. 11235003.

\end{document}